\documentclass[aps,prl,floatfix,twocolumn,footinbib]{revtex4}
\usepackage{epsfig}
\begin{document}

\title{The Ebers-Moll model for magnetic bipolar transistors}

\author{Jaroslav Fabian\footnote{email:jaroslav.fabian@uni-graz.at}} 
\affiliation{Institute of Physics, Karl-Franzens University, 
Universit\"atsplatz 5, 8010 Graz, Austria} 

\author{Igor \v{Z}uti\'{c}\footnote{email:zutic@dave.nrl.navy.mil}}
\affiliation{Center for Computational Materials Science,
Naval Research Laboratory, Washington, D.C. 20375 and
Condensed Matter Theory Center, Department of Physics, 
University of Maryland at College
Park, College Park, Maryland 20742-4111}

\begin{abstract}
The equivalent electrical circuit of the Ebers-Moll type is introduced
for magnetic bipolar transistors. In addition to conventional diodes and current
sources, the new circuit comprises two novel elements due to spin-charge coupling.
A classification scheme of the operating modes of magnetic bipolar
transistors in the low bias regime is presented.
\end{abstract}
\maketitle

Semiconductor spintronics \cite{Zutic2004:RMP} offers novel 
functionalities by combining electronics for signal processing and 
magnetism for both nonvolatility and additional
electronic control. At the present stage, with the fundamentals
of spin injection \cite{Fiederling1999:N, Young2002:APL, Jonker2000:PRB,Jiang2003:PRL}, 
spin relaxation \cite{Meier:1984, Kikkawa1998:PRL, Dzhioev2002:PRB}, 
as well as semiconductor magnetism \cite{Dietl2002:SST, Ohno1998:S}
established, there is a need for new ideas demonstrating practical
use of the fundamental spin physics.  
It was shown in Refs. \cite{Fabian2002:P, Fabian2004:APL, Fabian2004:PRB}
that magnetic bipolar transistors (MBT), which can employ both
ferromagnetic and paramagnetic semiconductors (with large g-factor) 
\cite{Zutic2004:RMP}, 
can significantly extend 
functionalities of conventional bipolar junction
transistors (BJT) \cite{Shockley1951:PR} by exploiting spin-charge coupling
of the Silsbee-Johnson type 
\cite{Silsbee1980:BMR, Johnson1985:PRL}. Current amplification in MBT's,
for example, can be modulated by magnetic field during the device operation,
giving rise to the phenomena of (giant) magnetoamplification 
\cite{Fabian2004:PRB}. 

In this paper we generalize the widely used Ebers-Moll equivalent circuit 
of BJT
\cite{Ebers1954:PIRE} 
(reprinted in \cite{Sze:1991}) to MBT. Two novel
electronic elements are added to the original circuit---spin
diodes and spin current sources---to describe spin-charge coupling. Our goal is
to provide a simple computational scheme for MBT's as well as
to show an example how a novel spintronics device
can be described by (and integrated with) a more conventional 
electronic circuitry.

\begin{figure}
\centerline{\psfig{file=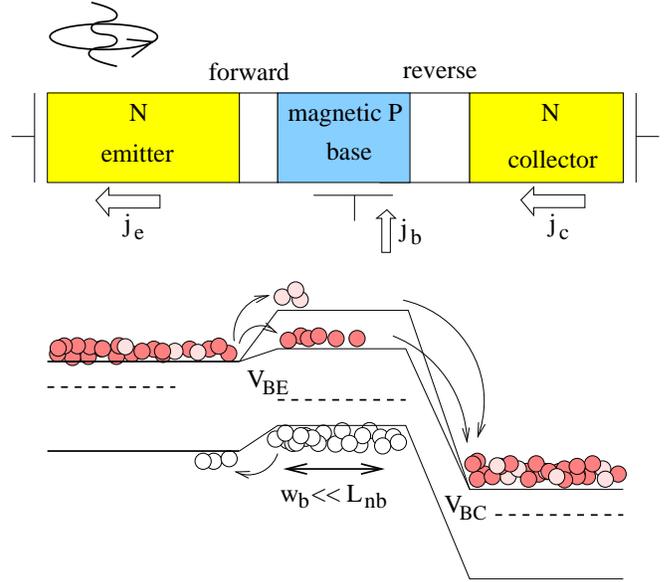,width=1\linewidth}}
\caption{Scheme of a {\it npn} magnetic bipolar transistor
in the forward active mode. The base has an equilibrium
electron spin polarization $P_{0b}$, illustrated by the spin-split
conduction band. Spin up (down) electrons are pictured as dark (light)
filled circles. Holes are unpolarized. The emitter has a source of
spin polarization, here shown as a circularly polarized light, giving
rise to a nonequilibrium spin polarization $\delta P_e$. The direction
of the currents is indicated. 
}
\label{fig:1}
\end{figure}

MBT's comprise two magnetic p-n junctions \cite{Zutic2002:PRL, Fabian2002:PRB} in series.
In the scheme of Fig. \ref{fig:1} we show an $npn$ MBT with a magnetic
base. Magnetic here means that there is an equilibrium spin splitting
$2q\zeta_b$ of the conduction band (valence band would also work), giving rise
to an equilibrium spin polarization $P_{0b}=\tanh(q\zeta_b/k_BT)$ in the base 
($T$ is temperature). Hole spins are assumed unpolarized.
Either a ferromagnetic semiconductor or a
diluted magnetic semiconductor with a large g-factor in a magnetic
field will work. Both the magnitude and the sign of $P_{0b}$ can 
be controlled by an external magnetic field (this is what we
call magnetic control). In addition
to the equilibrium spin, there can be a nonequilibrium (excess) spin
injected by external means (providing spin control) into the
emitter and collector. The corresponding spin 
polarizations are $\delta P_e$ and $\delta P_c$. 
If the bias on the base-emitter ($be$) and base-collector
($bc$) junction is $V_{be}$ and $V_{bc}$, respectively, then the 
excess electron densities in the base, close to the $be$ and $bc$ 
junctions, are \cite{Fabian2004:PRB}
\begin{eqnarray} \label{eq:dnbem}
\delta n_{be} & = & n_{0b}(\zeta_b)\left [ e^{qV_{be}/k_B T}
\left ( 1 + \delta P_e P_{0b} \right )-1 \right ],   \\
\delta n_{bc} & = & n_{0b}(\zeta_b) \left [ e^{qV_{bc}/k_B T}
\left ( 1 + \delta P_c P_{0b} \right ) - 1
\right ].
\end{eqnarray}
The influence of the equilibrium spin is felt both by the
equilibrium number of electrons in the base,  $n_{0b}(\zeta_b)=
n_{0b}(0)\cosh(q\zeta_b/k_BT)$, as well as by the spin-charge coupling
factor $1+\delta P P_{0b}$. The nonequilibrium spin plays a role only in the latter.
The unpolarized hole excess densities 
in the emitter and collector, close to the depletion layer with the base,  
are given by the standard formulas
\begin{eqnarray} \label{eq:dpe}
\delta p_{e} &=& p_{0e} (e^{qV_{be}/k_B T}-1), \\
\label{eq:dpc}
\delta p_{c} &=& p_{0c} (e^{qV_{bc}/k_B T}-1),
\end{eqnarray}
where $p_{0e}$ and $p_{0c}$ is the equilibrium number of holes
in the emitter and collector.

Electrical currents in MBT's can be expressed, as in BJT's
through the excess densities $\delta n_{be}$
and $\delta n_{bc}$ of the minority carriers\cite{Fabian2004:PRB}:
\begin{eqnarray} \label{eq:je}
j_e&=&j_{gb}^n\left [\frac{\delta n_{be}}{n_{0b}}-\frac{1}{\cosh(w_b/L_{nb})}
\frac{\delta n_{bc}}{n_{0b}} \right ] + j_{ge}^p\frac{\delta p_{eb}}{p_{0e}},\\
\label{eq:jc}
j_c&=&j_{gb}^n\left [-\frac{\delta n_{bc}}{n_{0b}}+\frac{1}{\cosh(w_b/L_{nb})}
\frac{\delta n_{be}}{n_{0b}} \right ]- j_{gc}^p\frac{\delta p_{cb}}{p_{0c}}.
\end{eqnarray}
The base current is $j_b=j_e-j_c$ and the electron generation current in the base is
\begin{equation} \label{eq:jgn}
j_{gb}^n=\frac{qD_{nb}}{L_{nb}}n_{0b}\coth\left(\frac{w_b}{L_{nb}}\right).
\end{equation}
Here $D_{nb}$ stands for the electron diffusion coefficient in the base
whose effective width is $w_b$ and  $L_{nb}$ is the electron diffusion length in the 
base (see Fig. \ref{fig:1}). The hole generation currents in the emitter, $j_{ge}^p$,
and collector, $j_{gc}^p$, are given similarly to Eq. (\ref{eq:jgn}) with
$n$ replaced by $p$ and $e$ replaced by either $e$ or $c$.

The Ebers-Moll model \cite{Ebers1954:PIRE} is an equivalent circuit 
to a BJT. We will now introduce this standard 
model and generalize it  to the case of MBT's.  
Denote by $j_{se}$ and $j_{sc}$ the emitter and collector
saturation currents (note that $s$ stands for saturation, not spin):
\begin{eqnarray}
j_{se}&=&j_{gb}^n+j_{ge}^p, \\
j_{sc}&=&j_{gb}^n+j_{gc}^p. 
\end{eqnarray}
The emitter saturation current is the emitter current that flows 
if $V_{be}<0$, $V_{bc}=0$, and only equilibrium spin present 
[$j_e=-j_{se}$, see Eq. (\ref{eq:je})]. 
Similarly for the collector saturation current.
Denote next the forward and reverse currents (terminology
from the forward active mode) as
\begin{eqnarray}
j_f&=&j_{se}(e^{qV_{be}/k_BT}-1), \\
j_r&=&j_{sc}(e^{qV_{bc}/k_BT}-1).
\end{eqnarray}
Finally, we introduce the spin-charge forward and reverse currents 
\begin{eqnarray} \label{eq:jmf}
j_{mf}&=&j_{gb}^n\delta P_e P_{0b}e^{qV_{be}/k_BT}, \\
\label{eq:jmr}
j_{mr}&=&j_{gb}^n\delta P_c P_{0b}e^{qV_{bc}/k_BT}.
\end{eqnarray}
Here subscript $m$ stands for magnetic to stress that the current,
that is due to spin-charge coupling across the depletion regions, 
appears only in magnetic transistors.
These currents flow due to the presence of nonequilibrium spin 
polarization and are finite even at zero bias 
(spin-voltaic effect\cite{Zutic2002:PRL}).

\begin{figure}
\centerline{\psfig{file=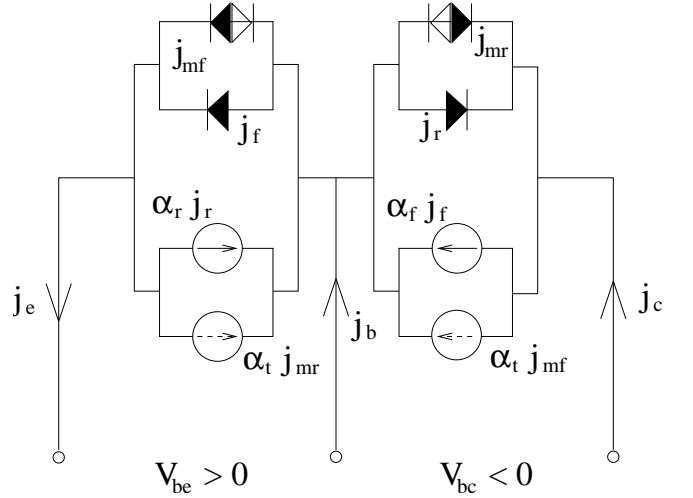,width=1\linewidth}}
\caption{The Ebers-Moll equivalent circuit
of a MBT. The voltage sources are arranged for the forward
active mode. The left (right) circuit is the emitter (collector).
The emitter circuit has a diode for the forward current,
and a current source which depends on the bias in the
collector circuit. In addition, there are two new elements. A spin diode
whose direction can be flipped: its filled triangle points to
the forward direction when $\delta P_{e} P_{0b} >0$, otherwise
the current direction changes.
A new spin current source (dashed arrow), pointing in the direction
of the current if $\delta P_{e} P_{0b} >0$. The direction
of the current can also be flipped.  Similar notation applies
for the right (collector) circuit.
}
\label{fig:EM}
\end{figure}

The generalized Ebers-Moll model directly derives from Eqs. (\ref{eq:je}) and
(\ref{eq:jc}), and reads 
\begin{eqnarray} \label{eq:jeem}
j_e&=&j_f - \alpha_r j_r + j_{mf} - \alpha_t j_{mr}, \\
\label{eq:jcem}
j_c&=& \alpha_f j_f - j_r +\alpha_t j_{mf} -j_{mr}.  
\end{eqnarray}
Here $\alpha_f$ has the meaning of the transport factor
in the forward active mode, while $\alpha_r$ is the transport factor in
the reverse active mode in the absence of spin-charge coupling, as can be 
seen directly from Eqs. (\ref{eq:jeem}) and (\ref{eq:jcem}). The transport
factor $\alpha_t$ is $\alpha_t=1/\cosh(w_b/L_{nb})$.
The conventional
Ebers-Moll model is recovered by putting $j_{mf}=j_{mr}=0$. As in the 
conventional model, the following equality holds:
\begin{equation}
\alpha_fj_{se}=\alpha_rj_{sc}.
\end{equation}
This can be verified by requiring that
\begin{equation}
j_e(V_{be}=0, V_{bc}=V)=-j_c(V_{be}=V, V_{bc}=0),
\end{equation}
for $\delta P_e=\delta P_c=0$.
In our ideal case it is straightforward to show that 
\begin{equation}
\alpha_f j_{se}= \alpha_r j_{sc} = \alpha_t j_{gb}^n.
\end{equation}

The equivalent circuit to Eqs. (\ref{eq:jeem}) and (\ref{eq:jcem}) 
is shown in Fig. \ref{fig:EM}. The current
flow is the same as in Fig. \ref{fig:1}. Let us discuss the emitter circuit.
It consists
of four elements: (i) a conventional diode with the directional current $j_f$ that depends
on $V_{be}$, (ii) a conventional current source giving current $\alpha_r j_r$
that depends on $V_{bc}$ and on the transport factor $\alpha_r$ measuring
the amount of current injected into the emitter from the collector, 
(iii) a spin diode with the forward current $j_{mf}$, and
finally, (iv) a spin current source $\alpha_t j_{mr}$. 
The first two elements appear already in BJT's.
The spin diode (iii), which appears due to spin-charge coupling, works
similar to a diode in the sense that its current is rectified with
$j_{mf} \sim \exp(qV_{be}/k_B T)$. The crucial difference from conventional
diodes is that the direction of the current flow can be changed by 
changing the sign of $\delta P_e P_{0b}$, see Eq. (\ref{eq:jmf}). 
The symbol for the spin-charge
diode reflects this fact. The filled triangle shows the direction 
when $\delta P_e P_{0b}$ is positive. The new functionality of MBT's then
lies in the ability to switch or modify the spin diode during
its operation. There is, in addition, the spin current source (iv) that 
is due to the electron current from spin-charge coupling. The
current, injected into the base from the collector,
diffuses towards the emitter through the base (this is why the
transport factor $\alpha_t$ appears). The element is a current source because
it does not depend on the voltage drop (here $V_{be}$) across it. 
It is, however, a controlled current source, similar to (ii), because it can be controlled
by $V_{bc}$. Because it arises from spin-charge coupling, the
magnitude and direction of (iv)
can be controlled by spin and magnetic field, adding to the
functionality of (iii).  Similar description applies to the collector circuit.  

\begin{table}
\begin{tabular}{|c|c|c|c|c|}
\hline
mode & $V_{be}$ & $V_{bc}$  & BJT & MBT  \\
\hline
\hline
forward active & F & R & amplification & MA, GMA \\
\hline
reverse active & R & F & amplification & MA, GMA \\
\hline
saturation & F & F & ON & ON, GMA, SPSW \\
\hline
cutoff & R & R & OFF & OFF \\
\hline
spin-voltaic & 0 & 0 & OFF & SPSW \\
\hline
\end{tabular}
\caption{Operational modes of BJT's and MBT's. 
Forward (F) and reverse (R) bias means positive and
negative voltage, respectively. Symbols MA and GMA stand for magnetoamplification and
giant magnetoamplification, while ON and OFF are modes of small
and large resistance, respectively; SPSW stands for spin switch.
}
\label{tab:BT}
\end{table}

For completeness we summarize in Tab. \ref{tab:BT} the operating modes (for a textbook
discussion see, for example, Ref. \onlinecite{Dimitrijev:2000}) of both BJT's
and MBT's, described by the Ebers-Moll model. Conventional
transistors have four modes, with amplification only in the forward and reverse
active modes (due to design only the forward active mode has significant current
gain). The saturation and cutoff modes are used in logic circuits for ON and OFF
states, respectively. MBT's have a much richer structure. In the active modes
both magnetoamplification \cite{Fabian2002:P, Fabian2004:PRB, Lebedeva2003:JAP} 
due to the dependence of the saturation currents on the equilibrium spin 
polarization and giant magnetoamplification 
\cite{Fabian2004:PRB} due to spin-charge coupling appear. In contrast to
conventional transistors, MBT's provide current gain even in the saturation
mode, due to spin-charge coupling. Furthermore, the transistor can act
as a spin switch, switching the current direction by flipping the spin 
\cite{Fabian2004:P}. In the cutoff mode MBT's are OFF and spin effects are
inhibited [see Eqs. (\ref{eq:jmf}) and (\ref{eq:jmr})]. Finally, a qualitatively
new mode, spin-voltaic, appears, due to spin-charge coupling. In this
mode, with no applied biases, the currents that flow are due only 
to the presence of nonequilibrium spin (which provide spin emf) 
and MBT's act as spin switches. 

In summary, we have generalized the Ebers-Moll model to 
include spin-charge coupling and cover magnetic bipolar transistors.
We have classified different operating modes of the transistors. In
most modes MBT's offer new functionalities such as spin switches or
magnetoamplifiers, which may have potential for signal processing,
logic circuits and nonvolatile memories. 

This work was supported by the US ONR, NSF, and DARPA. I. \v{Z}.
acknowledges financial support from the National Research Council.

\bibliography{references}

\newpage

\end{document}